\documentclass[sigplan,screen, noacm, authorversion, nonacm]{acmart}

\usepackage{mathtools}

\usepackage[disable]{todonotes}
\usepackage{amsmath,amsfonts}
\usepackage{graphicx}
\usepackage{xcolor}

\usepackage{booktabs}   
\usepackage{subcaption} 

\usepackage[frozencache,cachedir=.]{minted}
\usepackage[capitalise]{cleveref}
\usepackage{relsize}
\usepackage{paralist}
\usepackage{siunitx}
\usepackage{tikz}
\usepackage{xcolor}
\usepackage[all]{nowidow}

\usepackage{multirow}
\usepackage{tabularx}

\usepackage{pgfplots}

\usepackage{hyperref}

\usetikzlibrary{calc}
\usetikzlibrary{patterns,patterns.meta}
\usetikzlibrary{positioning}
\usetikzlibrary{shapes}
\usetikzlibrary{decorations,decorations.pathmorphing}
\usetikzlibrary{fit}
\usetikzlibrary{decorations.pathreplacing}
\usetikzlibrary{pgfplots.groupplots}
\usetikzlibrary{matrix}
\usetikzlibrary{arrows, chains, shapes.geometric, shapes.symbols}

\usepackage{fontawesome5}

\usepackage[all]{nowidow}
\usepackage{lipsum}  
\setlipsum{%
  par-before = \begingroup\color{red},
  par-after = \endgroup
}

\makeatletter
\newcommand{\removelatexerror}{\let\@latex@error\@gobble}
\makeatother

\usepackage[linesnumbered,vlined]{algorithm2e}
\usepackage{setspace}

\usetikzlibrary{calc,trees,positioning,arrows,chains,shapes.geometric,%
    decorations.pathreplacing,decorations.pathmorphing,shapes,%
    matrix,shapes.symbols,patterns}

\definecolor{col0o}{RGB}{0,146,146}
\colorlet{col0}{col0o!80}
\definecolor{col1o}{RGB}{182,109,255}
\colorlet{col1}{col1o!70}
\definecolor{col2}{RGB}{219,209,0}
\definecolor{col3}{RGB}{255,182,119}
\definecolor{col8}{RGB}{182,255,219}
\definecolor{col5}{RGB}{36,255,36}
\definecolor{col6}{RGB}{182,219,255}
\definecolor{col7}{RGB}{255,109,182}
\definecolor{col18}{RGB}{255,255,109}
\definecolor{col9}{RGB}{73,0,146}
\definecolor{col10}{RGB}{146,73,0}
\definecolor{col11}{RGB}{146,0,0}
\definecolor{col12}{RGB}{0,109,219}
\definecolor{col13}{RGB}{0,73,73}
\definecolor{col14}{RGB}{73,0,73}
\definecolor{col15}{RGB}{73,73,0}
\definecolor{col16o}{RGB}{255,36,36}
\colorlet{col16}{col16o!80}
\definecolor{col17o}{RGB}{36,36,255}
\colorlet{col17}{col17o!50}
\definecolor{col4}{RGB}{255,182,219}
\definecolor{col19}{RGB}{109,182,255}
\colorlet{commentcolor}{black!80}
\definecolor{maroon}{HTML}{990000}

\pgfplotscreateplotcyclelist{mycolorlist}{%
fill=col0\\%
fill=col1\\%
fill=col2\\%
fill=col11\\%
fill=col4\\%
fill=col13\\%
fill=col7\\%
fill=col12\\%
fill=col9\\%
fill=col10\\%
fill=col3\\%
fill=col5\\%
fill=col14\\%
fill=col15\\%
fill=col16\\%
fill=col17\\%
}
\colorlet{commentcolor}{black!80}
\colorlet{pragmacolor}{col10}

\usepackage{listings}
\usepackage{microtype}
\pgfplotsset{compat=1.18} 

\tikzset{
  >=stealth,
  thread/.style={->,decorate,
    decoration={snake,amplitude=.4mm, segment length=2mm, post length=1mm},
    thick},
  workerthread/.style={
    black!85,
    dash pattern=on 3pt off .5pt,
    thread,
  },
  sm/.style={
    rectangle,
    rounded corners,
    fill=white,
    draw=black, very thick,
    inner sep = 2mm,
    minimum width=3em,
    minimum height=3em,
    text centered},
  mynode/.style={
    rectangle,
    rounded corners,
    draw=black, very thick,
    inner sep = 2mm,
    minimum height=3em,
    text centered},
  line/.style={draw, thick, <-},
  element/.style={
    tape,
    top color=white,
    bottom color=blue!50!black!60!,
    minimum width=8em,
    draw=blue!40!black!90, very thick,
    text width=10em,
    minimum height=3.5em,
    text centered},
  every join/.style={->, ultra thick,shorten >=1pt},
  tuborg/.style={decorate, ultra thick,white,text=black},
  midnode/.style={midway, above=2pt},
  midnodebelow/.style={midway, below=2pt},
  mybrace/.style={decorate,decoration={brace,aspect=#1,amplitude=10pt}},
  device/.style={
    rectangle,
    rounded corners,
    fill=white,
    draw=black, very thick,
    inner sep = 2mm,
    minimum width=3.5em,
    minimum height=4em,
    text centered},
  host/.style={
    rectangle,
    fill=white,
    draw=black, thick,
    inner sep = 1mm,
    minimum width=1em,
    minimum height=1em,
    text centered},
}

\newdimen\XCoordA
\newdimen\YCoordA
\newdimen\XCoordB
\newdimen\YCoordB
\newdimen\XCoordC
\newdimen\YCoordC
\newdimen\XCoordD
\newdimen\YCoordD
\newdimen\XCoordE
\newdimen\YCoordE

\newcommand*{\ExtractCoordinateA}[1]{\path (#1); \pgfgetlastxy{\XCoordA}{\YCoordA};}%
\newcommand*{\ExtractCoordinateB}[1]{\path (#1); \pgfgetlastxy{\XCoordB}{\YCoordB};}%
\newcommand*{\ExtractCoordinateC}[1]{\path (#1); \pgfgetlastxy{\XCoordC}{\YCoordC};}%
\newcommand*{\ExtractCoordinateD}[1]{\path (#1); \pgfgetlastxy{\XCoordD}{\YCoordD};}%
\newcommand*{\ExtractCoordinateE}[1]{\path (#1); \pgfgetlastxy{\XCoordE}{\YCoordE};}%

\setminted{fontsize=\small}

\lstdefinestyle{c_code_style}{
basicstyle=\small\ttfamily,
keywordstyle=\color{blue},
commentstyle=\color{gray},
breaklines=true,
language=C,
morekeywords={uint64_t, int32_t, uint32_t},
}
\usepackage{enumitem}

\begin{document}

\title{Input-Gen: Guided Generation of Stateful Inputs for Testing, Tuning, and Training}


\author{Ivan R. Ivanov}
\orcid{0000-0003-0356-3768}
\affiliation{
  \institution{Tokyo Institute of Technology \\ RIKEN CCS}
  \city{Kobe}
  \country{Japan}
}
\email{ivanov.i.aa@m.titech.ac.jp}

\author{Joachim Meyer\footnotemark[1]}
\orcid{0000-0003-2656-9863}
\affiliation{
  \institution{Compiler Design Lab, Saarland University}
  \city{Saarbrücken}
  \country{Germany} 
}
\email{jmeyer@cs.uni-saarland.de}

\author{Aiden Grossman}
\orcid{0000-0001-9430-4782}
\affiliation{
  \institution{University of California, Davis}
  \city{Davis}
  \country{USA}
}
\email{amgrossman@ucdavis.edu}

\author{William S. Moses}
\orcid{0000-0003-2627-0642}
\affiliation{
  \institution{\hspace*{-1em}University of Illinois Urbana Champaign\hspace*{-1em}\\ Google Deepmind}
  \city{Illinois}
  \country{USA}
}
\email{wsmoses@illinois.edu}

\author{Johannes Doerfert}
\orcid{0000-0001-7870-8963}
\affiliation{
   \institution{Lawrence Livermore National Laboratory}
   \city{Livermore}
   \country{USA}
}
\email{jdoerfert@llnl.gov}

\renewcommand{\shortauthors}{Ivanov et al.}

\begin{abstract}

The size and complexity of software applications is increasing at an accelerating pace.
Source code repositories (along with their dependencies) require vast amounts of labor to keep them tested, maintained, and up to date.
As the discipline now begins to also incorporate automatically generated programs, automation in testing and tuning is required to keep up with the pace -- let alone reduce the present level of complexity.
While machine learning has been used to understand and generate code in various contexts, machine learning models themselves are trained almost exclusively on static code without inputs, traces, or other execution time information.
This lack of training data limits the ability of these models to understand real-world problems in software.

In this work we show that inputs, like code, can be generated automatically at scale.
Our generated inputs are stateful, and appear to faithfully reproduce the arbitrary data structures and system calls required to rerun a program function. By building our tool within the compiler, it both can be applied to arbitrary programming languages and architectures and can leverage static analysis and transformations for improved performance. 
Our approach is able to produce valid inputs, including initial memory states, for $90\%$ of the ComPile dataset modules we explored, for a total of $21.4$ million executable functions. 
Further, we find that a single generated input results in an average block coverage of $37\%$, whereas guided generation of five inputs improves it to $45\%$.

\end{abstract}

\maketitle
\def\thefootnote{*}\footnotetext{Work conducted while at Lawrence Livermore National Laboratory.}\def\thefootnote{\arabic{footnote}}
\renewcommand{\paragraph}[1]{\textit{\textbf{#1.}}}

\label{sec:introduction}
\section{Introduction}

Machine learning (ML) has made data one of the most valuable commodities.
New datasets are released every month to train and tune ever growing models.
While code datasets come in all shapes and forms~\cite{dasilve2021anghabench,kocetkov2022stackv1,lozhkov2024stackv2,grossman2023compile},
only a precious few provide means to execute the code~\cite{armengol2022exebench,berezov2022cola,kind2022jotai}.
Even performance benchmark suites, which are executable by design, tend to only offer a few end-to-end inputs.
Thus, ML training is heavily lopsided towards static code rather than dynamic code execution.

Program execution has always been a crucial part of testing and tuning efforts, even before machine learning.
The underlying issues in both testing and tuning are very much alike. Moreover, there is no versatile way to 
generate stateful inputs for an arbitrary program fragment. Without such inputs, we are reliant on
manual efforts or existing end-to-end inputs, even if most of the execution trace is irrelevant to the task at hand.
While prior art can record a trace of a given program execution~\cite{lee2005codeisolator,liao2010sourceoutlining,akel2013sourceisolation,castro2015cere,popov2015pcere,parasyris2023kernelrecordreplay}, the resulting profile coverage is limited by the diversity of the recorded traces.
Further, the initial recording needs an end-to-end run of the application, and, depending on the framework, the recording might not allow to replay parts of the program in isolation.

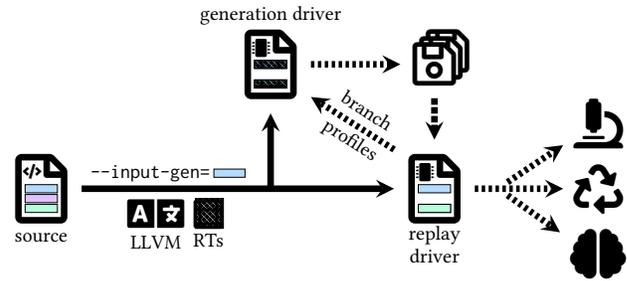
\begin{figure}
    \centering
    \resizebox{1.0\linewidth}{!}{\begin{tikzpicture}
  \tikzstyle{every node}=[font=\huge]
  \begin{scope}[local bounding box=system]

  \node[scale=3.5] (code) at (0.0,0.0) {\huge\faFile[]{}};
  \node[] (code_txt) at ($(code.south) + (0, 0.0)$) {source};
  \node[scale=0.9,anchor=south east] (code_sym) at ($(code) + (0.12, 0.2)$) {\LARGE\faCode[]{}};
  \coordinate (S) at ($(code.west)!0.6!(code)$);
  \coordinate (P) at ($(code)!0.15!(code.north)$);
  \coordinate (Q) at ($(code)!0.4!(code.east)$);
  \ExtractCoordinateA{S}
  \ExtractCoordinateB{P}
  \ExtractCoordinateC{code}
  \ExtractCoordinateD{Q}
  \draw[fill=col6,thick] ($(\XCoordD, \YCoordB) + (0, -0.10)$) rectangle ($(\XCoordA, \YCoordB) + (0, -0.35)$);
  \draw[fill=col1!60!white,thick] ($(\XCoordD, \YCoordB) + (0, -0.40)$) rectangle ($(\XCoordA, \YCoordB) + (0, -0.65)$);
  \draw[fill=col8,thick] ($(\XCoordD, \YCoordB) + (0, -0.70)$) rectangle ($(\XCoordA, \YCoordB) + (0, -0.95)$);
  

  \node[scale=3.5] (driver) at (7.0, 3.8) {\huge\faFile[]{}};
  \node[scale=3.5] (runner) at (12.0, 0) {\huge\faFile[]{}};
  \coordinate (S) at ($(driver.south)!0.5!(runner.south)$);
  \node[] (driver_txt) at ($(driver.north) + (0, 0.0)$) {generation driver};
  \node[text width=2cm, align=center] (runner_txt) at ($(runner.south) + (0,-0.2)$) {replay driver};
  \node[scale=1.1,anchor=south east] (driver_sym) at ($(driver) + (0.125, 0.1)$) {\LARGE\faMicrochip[]{}};
  \node[scale=1.1,anchor=south east] (runner_sym) at ($(runner) + (0.125, 0.1)$) {\LARGE\faMicrochip[]{}};
  \coordinate (S) at ($(driver.west)!0.6!(driver)$);
  \coordinate (P) at ($(driver)!0.15!(driver.north)$);
  \coordinate (Q) at ($(driver)!0.4!(driver.east)$);
  \ExtractCoordinateA{S}
  \ExtractCoordinateB{P}
  \ExtractCoordinateC{driver}
  \ExtractCoordinateD{Q}
  \draw[fill=col6,thick] ($(\XCoordD, \YCoordB) + (0, -0.10)$) rectangle ($(\XCoordA, \YCoordB) + (0, -0.35)$);
  \coordinate (exec1) at ($(\XCoordD, \YCoordB) + (0, -0.1)$);
  \coordinate (exec2) at ($(\XCoordD, \YCoordB) + (0, -0.35)$);
  \coordinate (exec_ep) at ($(exec1)!0.5!(exec2)$);
  \draw[pattern={Lines[angle=45,distance=1.5pt,line width=2pt]}] ($(\XCoordD, \YCoordB) + (0, -0.10)$) rectangle ($(\XCoordA, \YCoordB) + (0, -0.35)$);
  
  \draw[fill=col8,thick] ($(\XCoordD, \YCoordB) + (0, -0.70)$) rectangle ($(\XCoordA, \YCoordB) + (0, -0.95)$);
  \draw[pattern={Lines[angle=45,distance=1.5pt,line width=2pt]}] ($(\XCoordD, \YCoordB) + (0, -0.70)$) rectangle ($(\XCoordA, \YCoordB) + (0, -0.95)$);  
  
  \coordinate (S) at ($(runner.west)!0.6!(runner)$); 
  \coordinate (P) at ($(runner)!0.15!(runner.north)$);
  \coordinate (Q) at ($(runner)!0.4!(runner.east)$);
  \ExtractCoordinateA{S}
  \ExtractCoordinateB{P}
  \ExtractCoordinateC{runner}
  \ExtractCoordinateD{Q}
  \draw[fill=col6,thick] ($(\XCoordD, \YCoordB) + (0, -0.10)$) rectangle ($(\XCoordA, \YCoordB) + (0, -0.35)$);
  \draw[fill=col8,thick] ($(\XCoordD, \YCoordB) + (0, -0.70)$) rectangle ($(\XCoordA, \YCoordB) + (0, -0.95)$);

  
  
  \draw[fill=col8,thick] ($(\XCoordD, \YCoordB) + (0, -0.70)$) rectangle ($(\XCoordA, \YCoordB) + (0, -0.95)$);
  

  \node[scale=2.75] (compiler) at (3.5,-0.8) {\LARGE\faLanguage{}};
  \node[] (compiler_txt) at ($(compiler.south) + (0, 0.0)$) {LLVM};
  \node[scale=2.035, anchor=east, yshift=0.2mm] (runtime) at ($(compiler.east) + (1.1, 0)$) {\LARGE\faSquare[]{}};
  \node[] (runtime_txt) at ($(runtime.south) + (0,-0.12)$) {RTs};

  \coordinate (S) at ($(compiler.north)!0.2!(compiler.south)$);
  \coordinate (P) at ($(compiler.south)!0.25!(compiler.north)$);
  \coordinate (Q) at ($(compiler)!0.7!(compiler.east)$);
  \ExtractCoordinateA{compiler.south}
  \ExtractCoordinateB{compiler.south east}
  \ExtractCoordinateC{S}
  \ExtractCoordinateD{P}
  \ExtractCoordinateE{Q}
  \draw[pattern={Lines[angle=45,distance=1.5pt,line width=2pt]}] ($(\XCoordB, \YCoordC)$) rectangle ($(\XCoordB, \YCoordD) + (\XCoordE, 0) - (\XCoordA, 0)$);k

  \coordinate (S) at ($(code.north east)!1.0!(code.east)$);
  \coordinate (P) at ($(runner.north west)!1.0!(runner.west)$);
  \ExtractCoordinateA{S}
  \ExtractCoordinateB{P}
  \draw[line width=2mm,->] (S) -- node[pos=0.22,above] (ig_arg) {\texttt{-{}-input-gen=}} ($(\XCoordB, \YCoordA)$);
  \draw[fill=col6,thick] ($(ig_arg.east) + (-0.10, 0.130)$) rectangle ($(ig_arg.east) + (0.85, -0.120)$);

  \ExtractCoordinateA{P}
  
  \ExtractCoordinateB{driver.south}
  \draw[line width=2mm, ->] ($((\XCoordB, \YCoordA)$) -- (driver.south);


  \node[scale=2.5] (testing) at (17.0,2.0) {\huge\faMicroscope[]{}};
  \node[scale=2.5] (tuning) at (17.0,0) {\huge\faRecycle[]{}};
  \node[scale=2.5] (training) at (17.0,-2.0) {\huge\faBrain[]{}};
  

  \ExtractCoordinateA{runner}
  \ExtractCoordinateB{driver.east}
  \node[scale=3.0] (new_input2) at ($(\XCoordA, \YCoordB) + (4mm, 4mm)$) {\LARGE\faSave[]{}};
  \draw[fill=white,draw=none] ($(new_input2.north west) - (7mm, 7mm)$) rectangle ($(new_input2.south east) - (7mm, 7mm)$);
  \node[scale=3.0] (new_input1) at ($(\XCoordA, \YCoordB) + (2mm, 2mm)$) {\LARGE\faSave[]{}};
  \draw[fill=white,draw=none] ($(new_input1.north west) - (7mm, 7mm)$) rectangle ($(new_input1.south east) - (7mm, 7mm)$);
  \node[scale=3.0] (new_input) at ($(\XCoordA, \YCoordB)$) {\LARGE\faSave[]{}};
  \draw[line width=2mm, dashed,->] ($(driver.east)$) -- ($(new_input.west)$);

  \draw[line width=2mm, dotted,->] ($(new_input.south)$) -- ($(runner.north)$);
  \draw[line width=2mm, dashed,->] ($(runner.east)$) -- ($(tuning.west)$);
  \draw[line width=2mm, dashed,->] ($(runner.east) + (10mm, 2mm)$) -- ($(testing.south west) + (0mm, 4mm)$);
  \draw[line width=2mm, dashed,->] ($(runner.east) + (10mm,-2mm)$) -- ($(training.north west) + (0mm,-4mm)$);
  
  \draw[line width=2mm, dashed,->] ($(runner.north west) + (0mm,-3mm)$) -- node[pos=0.4, rotate=-33, text width=2cm] {branch\\[0.5em]profiles} ($(driver.south east) + (0,5mm)$);
  
  \end{scope}

\end{tikzpicture}}
    \vspace{-5mm}
    
    \caption{
    Sketch of the input generation framework for a file (left) containing three functions (top, middle, bottom). 
    The user chooses to generate inputs for the top function, which results in an instrumented generation driver and a replay driver.
    Assuming the bottom function is called by the top one, it is instrumented as well and stays in both binaries.
    The middle function is not reached from the top function and consequently dropped.
    Each instrumented run generates an input file that can be executed by the replay driver.
    Optional profile feedback can guide the generation of new inputs.
    }
    \label{fig:intro}
    \vspace{-7mm}
\end{figure}

In this work we tackle the problem of input generation for arbitrary program parts without the need for any user provided inputs.
The initial task is defined as follows:
Given a function definition with no other requirements, generate inputs, including memory state, such that the function can be invoked and executed to completion.
Note that the definition explicitly allows for the function to call other function definitions in the translation unit, as well as external declarations.
We do not restrict control flow or memory accesses.
Finally, we do not limit ourselves to specific kinds of functions, e.g., GPU kernels, and we do not require systems support beyond standard C/C++ runtimes.

We developed a compiler instrumentation with a lightweight runtime system that is scalable and versatile enough to be deployed in most environments.
%
The overall design, sketched in \Cref{fig:intro}, consists of an LLVM compiler pass to instrument the code, and two runtimes, one for generation of inputs and one to facilitate the replay.
Through helper scripts we provide a fully automatic system that takes one or more LLVM-IR sources files, and generates as many inputs for the user selected functions as requested.
The inputs can then be executed through the replay driver to enable (continues) testing, performance tuning, and ML training.

The contributions of this work are as follows:
\begin{itemize}
    
    \item A framework to create stateful inputs for any programming language which compiles to LLVM. 

    \item Static and dynamic guidance for the input generation to improve branch coverage.
    Using compiler provided hints and branch profiling, the generation runtime will choose values that steer control flow towards under-explored areas.

   \item A case study on matrix multiplication to study time and memory implications of input generation and replay.

   \item A large scale study of our framework on the ComPile dataset.
   Out of $23.9$ million functions, we can instrument 99\% and successfully replay 90\%.
   The initial branch coverage of $37\%$ can be increased up to $45\%$ if 5 inputs are generated.
    
\end{itemize}

We will provide a high-level overview of our framework before we compare our approach to related work in \cref{sec:related_work}.
The main ideas are detailed in \cref{sec:approach}.
We end with a thorough evaluation in \cref{sec:evaluation} and closing thoughts in \cref{sec:future_work}.
\begin{figure*}[h]
\centering
\begin{tabular}{|c|c|c|}
\hline
\textbf{(a)} input program&\textbf{(b)} instrumented program&\textbf{(c)} replay program\\
\begin{minipage}[T]{0.3\linewidth}
\begin{minted}[fontsize=\footnotesize, escapeinside=||]{cpp}
|\;|
struct Node {
    struct Node* next;
    double value;
};

double sum(struct Node* list) {
    double result = 0.0;
    while (list) {
    
        result += list->value;
        
        list = list->next;
    }
    return result;
}
|\;|
\end{minted}
\end{minipage}&
\begin{minipage}[T]{0.3\linewidth}
\begin{minted}[fontsize=\footnotesize, escapeinside=||]{cpp}
|\;|
void |\color{maroon}{\textbf{\texttt{ig\_gen\_sum\_entry}}}|() {

    struct Node* list = |\color{maroon}{\textbf{\texttt{ig\_arg\_ptr}}}|();
    sum(list);
}
double sum(struct Node* list) {
    double result = 0.0;
    while (list) {
        |\color{maroon}{\textbf{\texttt{ig\_read\_double}}}|(&list->value);
        result += list->value;
        |\color{maroon}{\textbf{\texttt{ig\_read\_ptr}}}|(&list->next);
        list = list->next;
    }
    return result;
}
|\;|
\end{minted}
\end{minipage}&
\begin{minipage}[T]{0.3\linewidth}
\begin{minted}[fontsize=\footnotesize, escapeinside=||]{cpp}
|\;|
void |\color{maroon}{\textbf{\texttt{ig\_run\_sum\_entry}}}|() {
    |\color{maroon}{\textbf{\texttt{ig\_init\_memory}}}|();
    struct Node* list = |\color{maroon}{\textbf{\texttt{ig\_arg\_ptr}}}|();
    sum(list);
}
double sum(struct Node* list) {
    double result = 0.0;
    while (list) {
    
        result += list->value;
        
        list = list->next;
    }
    return result;
}
|\;|
\end{minted}
\end{minipage} \\
\hline
\end{tabular}
\vspace{-3mm}
    \caption{(a) depicts a simple program which sums up values in a linked list. (b) shows the same program, instrumented for input generation. This captures all "side-effects" necessary to reproduce the execution at a later time. Specifically, we indicate that we need a value of pointer type for the argument, and that we will load from \texttt{list->value} and \texttt{list->next}. (c) presents the version instrumented to run the generated input. Prior to calling the \texttt{sum()} function, the runtime will set up the appropriate memory state for execution and will provide the function argument.}
    \label{fig:linked_list_example}
\vspace{-2mm}
\end{figure*}
\section{High-Level Overview}
\label{sec:background}

Throughout this work, when we talk about the function for which input is generated we implicitly include all transitively reached function definitions in the same source file.
The goal is to create an input for a function, which consists of an initial memory state together with arguments and a sequence of values to be returned by calls to external functions.
Together, this input generally allows successful execution of the function.
Thus, calling the function with the input arguments in the input memory state will result in the execution being returned to the caller, or terminate through a routine like \texttt{exit}.

Examples, like the one in \cref{fig:linked_list_example}, are presented in C for readability, but our instrumentation pass operates on the LLVM-IR representation of the source code.
While it allows us to reuse existing compiler analyses and transformations, it is not a conceptual requirement.
Our runtimes are written in C++ with a very minimal C interface to simplify the communication with the instrumentation pass.

In general, we assume external symbols, including global and function declarations, are not available for input generation or replay.
However, we allow some known library functions, like \texttt{malloc}, as well as symbols related to (C++) error handling.

Memory is conceptually separated into two kinds, runtime managed objects and user managed allocations.
At the start of the generation and replay, there are no user managed allocations.
Only user code can create those via stack allocations, or calls to memory allocation routines we support natively, e.g., \texttt{malloc}.
Global memory, and other memory state the function assumes to be allocated prior to its execution, are managed by the runtime.
Global state is provided eagerly before the function execution starts, all other memory is allocated and initialized lazily as required.
The runtime tracks what memory regions are initialized, and records all values that have to be setup in memory, or returned from external functions during replay.
The interaction with the generation runtime is performed through a small set of runtime functions summarized in \cref{tbl:runtime_calls} which will be discussed in \cref{sec:approach}.

A sample execution of the instrumented program for input generation, shown in \cref{fig:linked_list_example} (b), could result in the following interaction:
1) The generation driver will setup the runtime and execute \texttt{ig\_gen\_sum\_entry()}, the entry point for the \texttt{sum} user function.
2) When \texttt{ig\_arg\_ptr()} is called, a new object is created and a pointer to it is returned.
3) The list traversal starts as the pointer was not null.
4) The runtime is informed that the \texttt{value} member of the list node is read. Since the location was unitialized and is runtime-managed, a double value is generated and stored there.
5) When the runtime is informed of the pointer read for the \texttt{next} member, it will create a new object, or return a null pointer, based on the result of a random value draw.
Assuming a new object is created, we restart the loop at 3).
Assuming a null pointer was returned, the loop is exited and the function finishes execution.
6) After completion of the function the memory state, that is all the objects that have been created together with the values that have been generated during the execution, are stored to enable  replay.

\section{Related Work}
\label{sec:related_work}

\paragraph{Input Recording}
Input recording involves techniques used to instrument a program to record inputs to specific code sections, typically referred to as codelets, extract these codelets, and then replay the recorded inputs against the codelets.
Use cases for this tooling can include debugging~\cite{ocallahan2016rr}, performance introspection~\cite{parasyris2023kernelrecordreplay,castro2015cere},
and more.
Input recording frameworks typically work by adding instrumentation at compile-time, with a notable exception being RR~\cite{ocallahan2016rr}, which does not require any compile-time instrumentation.
After possible instrumentation, the application under analysis is then run, producing a set of inputs that can be used to execute individual parts of the application.

However, the data dumps containing these inputs are often quite large.
The data dumps produced by CERE are tens to hundreds of megabytes due to the page-granularity capture of all memory touched during a function invocation.
Other work~\cite{parasyris2023kernelrecordreplay} dumps all allocations made by the application on the offloading device.
While these issues can be partially remedied through techniques like clustering~\cite{castro2015cere} or with more engineering work, such as the suggested tracking of deallocations in~\cite{parasyris2023kernelrecordreplay}, stored inputs still tend to be large.
More work~\cite{guo2008r2} has explored the use of custom annotations to steer where recording is performed to significantly reduce recording log size.
Reducing the captured inputs to a representative set allows for overall application performance to be extrapolated from fewer runs, which grants faster turn around times~\cite{castro2015cere,akel2013sourceisolation}.
In addition to size overhead, input recording frameworks can impose a large runtime overhead on apps being recorded, sometimes on the order of 100 times, precluding their use in some cases.


\paragraph{Input Generation}
Jotai~\cite{kind2022jotai} takes C files from open source repositories and attempts to make them executable through automatic derivation of parameter constraints. 
However, Jotai has several limitations around the types of inputs that it is able to generate, particularly requiring all functions that are called to be fully defined, leaving them with only 7\% of the dataset that they evaluated against, AnghaBench~\cite{dasilve2021anghabench}.
Jotai was only able to generate inputs for approximately 52\% of inputs matching the criteria.
Our work handles calls to functions without a definition by providing the function with a stubbed definition.

\paragraph{Benchmark Synthesis}
For the purpose of performance introspection at a variety of levels, a body of previous work has focused on \textit{synthesizing} benchmarks rather than trying to execute existing snippets of code.
Some~\cite{van2012synthesis} focus on synthesizing benchmark proxies for existing sets of code to hide proprietary information.
Genesys~\cite{panda2016genesys} tackles the problem by generating assembly-level snippets that match a variety of key instruction, control-flow, and memory-access patterns of a specific workload.
COLA-GEN~\cite{berezov2022cola} is designed to take a function specification in a custom DSL and output a stand-alone executable C program which matches those characteristics.
While these techniques can often create a large number of benchmarks, whether or not these benchmarks are faithful to real workloads beyond simple properties like instruction mix is currently an open question.
Some techniques, like the one used for Genesys~\cite{panda2016genesys} also preclude the use case of compiler testing by directly generating assembly rather than higher-level IR or C.

In addition to classical techniques for benchmark synthesis, there is also an increasingly wide body of work utilizing machine learning for the synthesis of benchmarks.
CLGen~\cite{cummins2017predictivemodeling} uses language modeling to synthesize OpenCL kernels, which the authors demonstrate to be similar to human-written kernels through the use of a Turing test, despite their relative simplicity.
In addition, CLGen is only able to generate benchmarks that compile 2.33\% of the time~\cite{tsimpourlas2022benchpress}.
Taking advantage of recent advancements in language modeling,
BenchPress~\cite{tsimpourlas2022benchpress,tsimpourlas2023benchdirect} is able to improve the number of compilable kernels generated from 2.33\% to 86\% along with increasing diversity of the generated kernels.
Other work, such as \texttt{LaSynth~\cite{chen2021latentexecution}}, can be used to generate code for specific input-output pairs.
It achieves reasonable accuracy, but only for relatively simple snippets (tens of tokens), and additionally requires input-output pairs to be available to generate code.

\paragraph{Executable Datasets}
In addition to input recording and generation, there has also been previous work on utilizing these technologies in addition to other techniques to create datasets of executable code.
Some datasets, like the CodeContests utilized in AlphaCode~\cite{li2022alphacode}, utilize self-contained programs from competitions
that are inherently executable.
However, competition datasets often contain only a few clusters of programs not necessairly representative of real world use cases.
Other datasets, such as ExeBench~\cite{armengol2022exebench} attempt to remedy this problem by synthesizing inputs for arbitrary C code pulled from Github and AnghaBench~\cite{dasilve2021anghabench}.
However, the methodology used in ExeBench is only able to generate input-output pairs that are runnable in approximately 15\% of cases.

\paragraph{Test Generation and Fuzzing}
Another body of work related to the input generation problem is automatic test generation.
There are several works~\cite{lemberger2021prtest,garg2013feedbackconcolic,pacheco2007feedbackrandomtest} that utilize feedback to help increase coverage of functions under test.
Several other pieces of work use of symbolic execution techniques~\cite{garg2013feedbackconcolic,visser2004pathfinder,yoshida2017klover,ringer2017solverlanguage} to increase coverage.
We leave integration of sophisticated symbolic execution techniques to increase input coverage in our domain to future work.
Recently, additional work has started exploring ML-guided test generation~\cite{liu2023chatgptcorrectcode,alshahwan2024unittest}, which has been shown to improve code coverage in large existing codebases.
In addition, fuzzing techniques often tackle a problem similar to our work for a different purpose, especially those aimed at compiler testing.
Tools like CSmith~\cite{yang20211csmith}
and
DeepSmith~\cite{cummins2018deepsmith}, generate executable programs intended to test compilers.
These programs are typically not representative of real world inputs, however, instead they are intended to exploit edge cases within the compiler.

\paragraph{Sanitizers}
Finally, another body of work related to the input generation problem are sanitizers.
Sanitizers can be used for detecting memory issues like use-after-frees and using uninitialized memory~\cite{stepanov2015msan,serebryany2012asan}, detecting threading issues like dead-locks and races~\cite{serebryany2012tsan}, and even detecting numerical accuracy issues~\cite{courbet2021nsan}.
Sanitizers typically consist of two main parts:
compile-time instrumentation, and a runtime, calls for which are inserted during the instrumentation phase.
We utilize a similar structure for our work.

\section{Approach}
\label{sec:approach}

Our approach is mainly comprised of 3 components: the instrumentation pass, a runtime for generation inputs, and a runtime for replaying them.
The compiler pass prepares the source module and introduces instrumentation that will instruct the runtime to generate stateful inputs during the program execution.
The replay runtime reads the generated stateful input description and sets up the memory state for replay and then invokes the
function in question.
We detail the various steps in the following subsections.

\subsection{Source Module Preparation}
\label{sec:preparation}

Module preparation comes in two variants, necessary and supportive.
The latter is mostly required to allow automation of our large scale studies across multiple programming languages that were not meant to be compiled with a vanilla LLVM compiler.
These transformations do not limit the generality of our approach but simplify automation.

\subsubsection{Linkage}
\label{sec:visibility_and_linkage}
In order to avoid conflicts of symbols found in the program with ones that may be linked in later, we prefix all symbols, including the user code, with a custom string in the compiler namespace (\texttt{\_\_ig...}).
For brevity, our examples do not show this step.
The renaming is not performed for a set of symbols that we want to share with the system, such as \texttt{malloc}, C++ exception handling functions, and special global variables, e.g., \texttt{stderr}.
To improve our static code analysis during the instrumentation for input generation we change the linkage of all symbols, except the generated entry points, to internal.
This is not strictly necessary but can improve the success rate for input generation.
For the replay runs we keep the original linkage and match the original compilation environment as no static analysis is performed.

\subsubsection{Function and Global Variable Declarations}
\label{sec:function-decls}
As shown in \cref{fig:function_stubs}, translation units are generally not self contained and instead contain reference to external functions and global variables defined elsewhere.
To compile such a module in isolation we must provide definitions for all declarations.
To this end, we replace each function declaration with a definition stub of the same type and name.
If the function has a non-void return, the stub will construct and return a new value of appropriate type.
These new values are tracked as any other value that is created during the generation (ref.~\cref{sec:value-gen}).
Global variables are defined as well and then processed with the existing global definitions.

\subsubsection{Debug Information}
\label{sec:debug-info}
Keeping debug information updated, especially when function declarations are defined, was not the aim of this work.
Consequently, we decided to drop all debug information present and thereby avoid problems that inconsistent debug
information can cause during code generation.



\subsection{Module Instrumentation}

To the outside, the generation runtime exposes a simple C interface that the compiler instrumentation pass targets. 
The families of available runtime calls are summarized in \cref{tbl:runtime_calls} and described below.

\begin{figure}[t]
\centering
\begin{minipage}{0.95\linewidth}
\begin{minted}[fontsize=\footnotesize, escapeinside=||]{cpp}
// Definitions are in other translation unit
// but type definitions are available.
extern configuration config;

bool read_database(Database *db);

bool open(Database *db) {
  if (config.read_db)
    return read_database(db);
  return create_database(db);
}
\end{minted}
\centering \textbf{(a) Code using function and global variable declarations.}
\end{minipage}\\[2em]

\begin{minipage}{0.95\linewidth}
\begin{minted}[fontsize=\footnotesize, escapeinside=||]{cpp}
static configuration |\color{maroon}{\textbf{\texttt{*config\_ptr}}}|;

static bool read_database(Database *db) {
  return |\color{maroon}{\textbf{\texttt{ig\_gen\_bool}}}|();
}

bool |\color{maroon}{\textbf{\texttt{ig\_gen\_open\_entry}}}|() {
  |\color{maroon}{\textbf{\texttt{ig\_reg\_global}}}|(&config_ptr, sizeof(configuration));
  Database *db = |\color{maroon}{\textbf{\texttt{ig\_arg\_ptr}}}|();
  return open(db);
}

bool open(Database *db) {
  // Use runtime managed global replacement
  |\color{maroon}{\textbf{\texttt{configuration &config = *config\_ptr}}}|;
  if (config.read_db)
    return read_database(db);
  return create_database(db);
}
\end{minted}
\centering \textbf{(b) Code after instrumentation and stub creation.}
\end{minipage}
\vspace{-3mm}
    \caption{
    Function and global variable declarations (top) that are replaced by definitions as part of the instrumentation pass (bottom).
    For global variables one level of indirection is introduced to allow their memory to be part of the runtime memory pool.
    The indirection is eliminated at the beginning of each function that uses the global.
    }
    \label{fig:function_stubs}
\end{figure}

\subsubsection{Generating Values}
\label{sec:value-gen}
We provide a family of entry points that can generate values of a specific type.
The API is used externally when a free standing value is needed, e.g., the return of an external function, or internally, when the program loads from not yet initialized memory.
There are two distinct kinds of values that can be generated, numeric values, e.g., integer and floating point values, and pointers.
For numeric values we default to querying a preset random distribution if no compiler provided hints are present.
The distribution is configurable at generation time and usually a standard random distribution.
For pointers, a new object is generated which is backed up by memory to allow future accesses. This will be further discussed in \cref{sec:objects}.
Compiler generated hints (ref.~\cref{sec:value-gen-hints}) will influence the decision, e.g., to favor a value for which a later branch condition will result in exploration of a non-traversed path.
Similarly, compiler and runtime feedback can influence the pointer choice and, for example, allow or disallow null pointers, or direct the generation runtime to pick a pointer into a specific existing object.

\subsubsection{Generating Arguments}
\label{sec:generating_arguments}
For the arguments to the function we generate new values of the appropriate type, as shown in \cref{fig:linked_list_example} (b), (c) and \cref{fig:function_stubs} (b).
Note that this utilizes the same mechanism to generate new values of a specific type described above.
However, the arguments are encoded separately in the stored input (ref.~\cref{sec:storeing_and_loading_inputs}).

\begin{figure}
\vspace{-3mm}
\scriptsize
\begin{tabular}{lll}
Action & Name & Call location \\
\hline
generating values & \texttt{ig\_gen\_<type>} & func. stubs \& internal\\
generate argument & \texttt{ig\_arg\_<type>} & entry points \\
memory read & \texttt{ig\_read\_<type>} & before memory reads\\
memory write & \texttt{ig\_write\_<type>} & before memory writes\\
register global & \texttt{ig\_reg\_global} & entry points \\
reg. branch conditions & \texttt{ig\_br\_cmp} & before branches \\
reg. pointer comparisons & \texttt{ig\_ptr\_cmp} & before pointer comp. \\
terminate execution & \texttt{ig\_exit} & ``unreachable'' locations \\
\end{tabular}
\vspace{-3mm}
\caption{We instrument the module for generation using the runtime C API. the \texttt{read}, \texttt{write}, \texttt{arg}, and \texttt{gen} runtime calls are defined for all primitive types.}
\label{tbl:runtime_calls}
\vspace{-3mm}
\end{figure}

\subsubsection{Accessing Memory}
\label{sec:memory}

Handling memory accesses is the cornerstone of the generation runtime.
Prior to each read and write access, the instrumentation inserts a call to inform the generation runtime of the accessed location and the type.
The memory being accessed is either runtime-managed, i.e., if it is within a memory region assigned to an object by the runtime, or not.
If it is not, that means it is either stack-allocated memory or was allocated by the user via allowed native routines, e.g., \texttt{malloc}.
Details on the object handling and initialization tracking are provided in \cref{sec:objects}.

\paragraph{Reading Memory}
The instrumentation will indicate a read access just before it happens in the program, as highlighted in \cref{fig:linked_list_example} {(b), (c)}.
For known access patterns, like a \texttt{memcpy}, the size is also included.
If the accessed location falls into an object managed by the runtime and the location has not been accessed before, we ensure that the program can actually read a value appropriate for the type.
This happens by generating a new value as described above. 
If the underlying memory is user-managed or has been accessed before, no action is taken.

\paragraph{Writing Memory}
Just prior to memory writes, the runtime is informed of the accessed location and the size of the write.
If the location is runtime-managed, we mark it as initialized but no other action is required.


%
%

\subsubsection{Registering Global Variables}
\label{sec:global-vars}
Global variables need to be registered with the runtime as they might have been initialized prior to the function execution in the original application.
The eager registration of all global variables is coupled with a level of indirection introduced by the compiler pass.
This is shown for the \texttt{config} variable in \cref{fig:function_stubs}.
The instrumentation provides the runtime with information about the global size and allows it to place them in the runtime-managed memory region next to other objects the user code assumes to be allocated prior to execution.
Accesses to global variables then simply reuse the logic described in above and detailed in \cref{sec:objects}.
For all global variables that are used in a function, loads at the beginning of the function will undo the indirection introduced by the instrumentation and allows the original code to execute as before.

\subsubsection{Aggregate Types}
\label{sec:aggregate}
Since aggregate types such as arrays, structs, or vectors can be arbitrary, we cannot provide runtime call for each possible type.
Instead, the instrumentation pass decompose aggregate types to their primitive type components and utilizes the associated runtime calls.
Complex types in all places of interaction between the program and the runtime are handled this way.


\subsubsection{Generation of Value Hints}
\label{sec:value-gen-hints}
To improve coverage of the code we are generating inputs for, we statically analyze which values generated by our
runtime influence control flow conditions. For branches that depend on a value that is to be generated,
we encode the condition that the value needs to satisfy for the different outcomes of the branch, e.g., that
the value has to smaller than 42. If profiling data is available, it is attached to the branch outcomes to
allow the runtime to choose a value that either satisfies the condition, or does not satisfy the condition,
depending on which branch result had lower coverage. The current implementation of branch hints is limited
to values generated close to the branch they influence but there is no fundamental reason the implementation could
not be enhanced in the future.

\subsubsection{Recording of Pointer Comparisons}
Similar to branch conditions, pointer comparisons are instrumented and reported to the runtime.
This allows introspection into the choice of pointer values that have been returned, and is used in the roll back scheme described in \cref{sec:rollback_offset}.

\subsubsection{Generation of Function Pointers}
\label{sec:indirect_calls}
Programs use function pointers to interact with shared libraries or to keep code generic.
While they are not immediately distinguishable from data pointers in LLVM-IR~\footnote{LLVM-IR uses ``opqaue'', or untyped pointers because pointer types in languages like C can be changed arbitrarily prior to an access.}, they require a conceptually different memory backing, namely executable code.
To this end, the instrumentation pass will first try to identify when any runtime generated value is used as the callee of an indirect function call.
A list of potential callees, that is functions in the module with a matching signature, is communicated to the runtime such that one of them can be picked at runtime.
To avoid inducing infinite recursion, we eliminate all functions than can reach the caller.
To account for the fact that function pointers may also originate from outside the module, and to ensure at least one callee is available, we add a function stub with a matching signature to the list of potential candidates.



\subsubsection{Graceful Program Termination}
\label{sec:program-exits}
Programs can contain locations that should not be reached, e.g., after a non-returning function like \texttt{exit} was called.
Assertions and error handling code are a common sources of such locations.
To avoid execution stopping before the generated input is stored, we replace non-returning calls with a runtime call which will trigger the runtime to tear down in an orderly fashion.

\begin{figure}[t]
\includegraphics[width=0.75\linewidth]{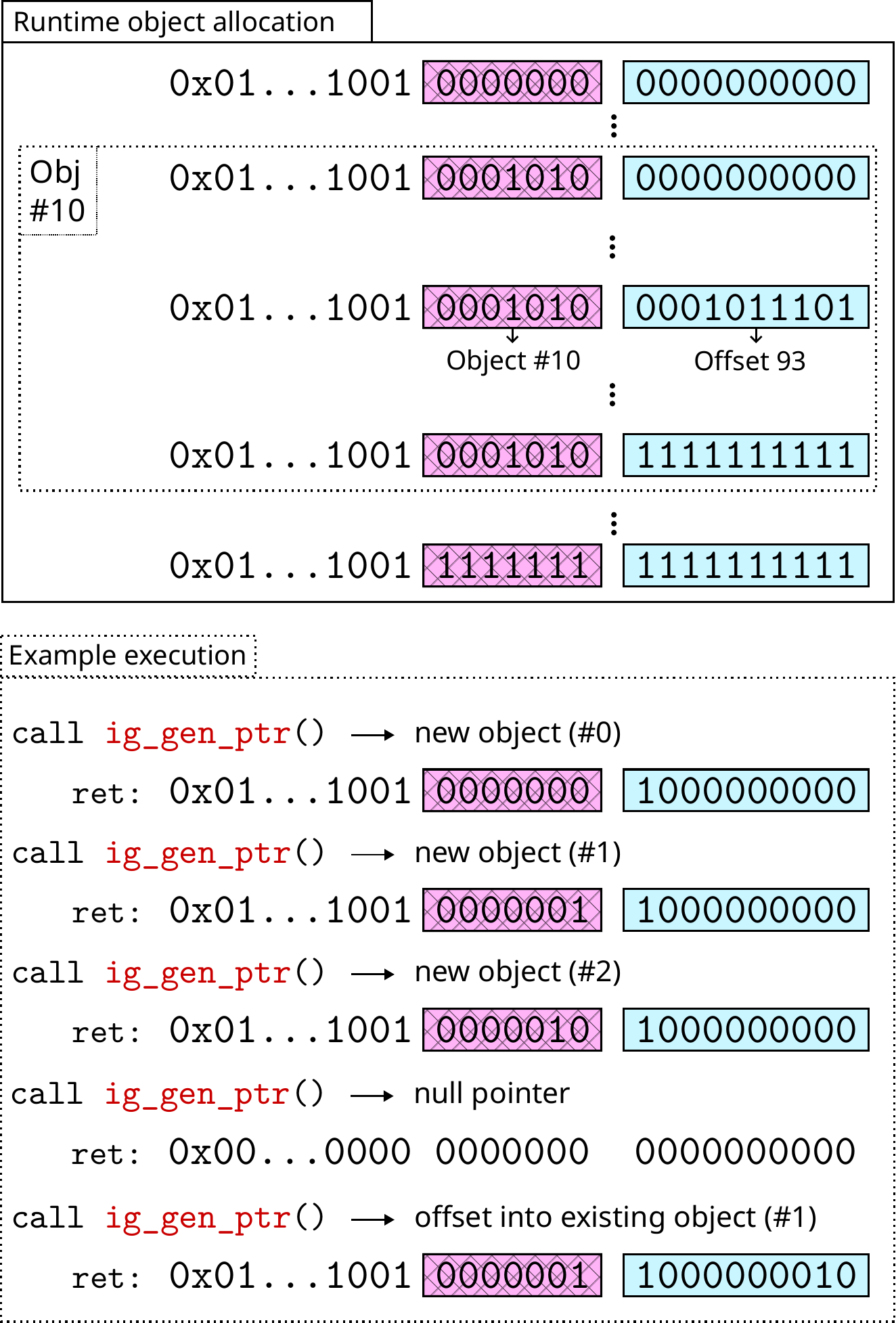}
\caption{In order to simplify and optimize object addressing, we allocate a large amount of memory for objects handled by our runtime, and implement a logical memory space partitioning strategy where the top bits are the index of the object itself, and the bottom bits refer to the offset in the specific object.  Note how when the runtime generates a pointer to a new object, it returns a pointer to the middle of the object, as the program may access objects at negative offsets as well.}
\label{fig:object_addressing}
\end{figure}

\subsection{Pointer Generation and \emph{Objects}}
\label{sec:objects}

The input generation runtime uses the notion of \emph{objects} to reason about memory accesses.
Objects are continuous memory regions that are accessed through offsets to a base pointer.
We assume there are only two kinds of memory accesses happening in the program.
First, accesses to memory allocated by the user function during execution, such as memory from \texttt{malloc} or stack-allocated memory.
We detect such accesses statically and dynamically, avoiding instrumentation in the first place if possible, and otherwise simply ignore them in the runtime.
All other accesses have to target objects through offsets of pointers that the runtime generated. 
These accesses are instrumented to allow objects to grow as necessary for the particular program run.
While hard-coded addresses, e.g., \texttt{*(0x8000) = ...}, are possible, our evaluation shows it is reasonable to ignore them for user-space code. There are no conceptual restrictions to allocating objects at hard-coded addresses, and thus support is possible
in the future.

All runtime-managed memory is part of a pre-allocated memory pool that is initialized eagerly.
Pool size, and other hyper-parameters of the described scheme can be adjusted as necessary.
%
%
For memory efficiency and runtime performance, we partition the memory pool and introduce virtual addressing.
The top bits of a pointer provided to the user code indicate the object ID, and the bottom bits indicate the offset within the object.
The scheme is visualized in \cref{fig:object_addressing}.
When a new pointer is requested by the program, the runtime can choose between three cases: a) Create a new object, b) use an offset into an existing object, or c) return a null pointer.

\subsubsection{Generating New Objects}
To generate a new object, we choose a region of unused memory in the pool and associate it with the object. We then return an aligned pointer\footnote{Many (vector) memory instructions need addresses to be aligned to a certain width.
We use 16 byte alignment for new objects by default.} to the middle of the region.
This allows the program to offset it both in the negative and positive direction while remaining in the region assigned to the object.
The maximal extent of the object is generally assumed to be unknown but the instrumentation pass provides static analysis results to minimize memory usage for global variables which are packed aggressively.
The maximal extent of objects of unknown size is limited by the number of bits we use to represent the offset in the virtual addressing scheme.

\subsubsection{Generating Offsets into Existing Objects}
\label{sec:rollback_offset}
Programs might rely on the fact that two pointers actually point into the same object, e.g., when traversing an array from the \texttt{begin()} pointer to the \texttt{end()} pointer.
For now, we handle this case via a rollback and retry strategy.
The compiler pass will instrument all pointer comparisons such that the runtime is aware of the pointers and objects involved.
Without prior knowledge, the runtime will always generate new objects when a fresh pointer value is requested.
If the instrumentation identifies a comparison between different objects, it is recorded as a constraint.
At this point we might choose to stop the generation process and roll back to the beginning.
In any subsequent re-run, the recorded constraints are used to emit offsets into existing objects rather than new ones.
The object that was created later is instead replaced with a offset into the object it was compared to.
The roll back is chance based and the offset is chosen to make the later comparison succeed.
This process is iterative and the constraints are accumulated until a valid input is created or the timeout is reached.
While static analysis can be used in the future to avoid roll backs, they will always be necessary for more complex examples in which there can be arbitrary computation between generating the first object being compared, generating the second object, and the actual comparison.

\begin{figure}[t]
\includegraphics[width=0.83\linewidth]{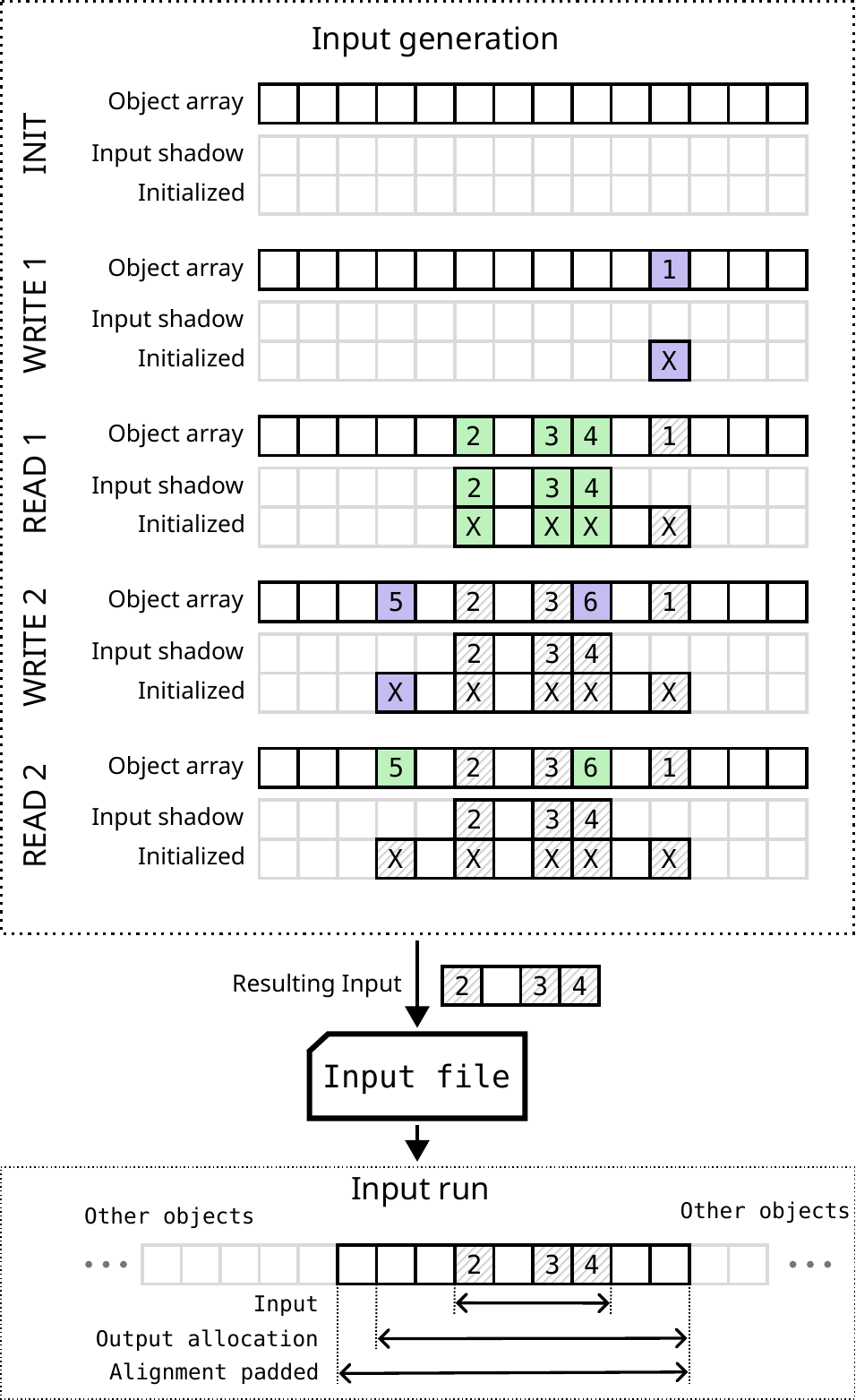}
\caption{The input generation runtime keeps track of the program memory (object array), input memory, and the parts of those that are used. We allocate memory on-demand for the input shadow and tracking initialization state (grayed-out vs solid cells). On the first read, we update both the program array and the input as it is a read from uninitialized memory. Writes only update the object array and the initialization state, so the input is kept intact, while reads from already initialized memory just return the value. At the end, only the input portion is saved to the binary file, while aligned space for the entire output is allocated when running.}
\label{fig:object_handling}
\end{figure}

\subsubsection{Generating Null Pointers}
\label{sec:rollback_null}
Some programs, such as the example in \cref{fig:linked_list_example}, expect a pointer to be null at some point in the execution.
We implement two strategies to deal with these cases.
The first one is to stochastically generate a null pointer when a fresh pointer is requested.
This, however, results in crashes and failed input generation in many cases where the pointer is expected to be valid and dereferenced.
The second approach involves the same strategy we use for handling offsets into objects.
In most cases, when a null pointer is expected somewhere, the pointer is eventually compared against a null constant.
Once such a comparison is reported to the runtime, a chance based rollback might be performed after which a null pointer replaces the object that was compared against null.


\subsection{Storing and Loading Inputs}
\label{sec:storeing_and_loading_inputs}
To facilitate replay, the generation runtime needs to store the generated values for later consumption.
We use a custom binary file format that contains the information required to reconstruct the memory state prior to the program execution, and replay all decisions.
Thus, the stored input file contains the generated objects with their partially initialized state, initial values for global variables and function arguments, and separately, the sequence of values that were generated by function stubs.

\paragraph{Tracking Objects and Minimal Initial State}
Consider the example in \cref{fig:object_handling} in which a sequence of memory reads and writes is performed on an initially unused object in runtime-managed memory (\texttt{INIT} rows).
In addition to the memory pool that holds the objects, here identified as \emph{Object array}, the runtime keeps two other arrays of correlated data.
The \emph{Input shadow} array keeps track of the initial memory state, thus the state of the memory prior to the execution of the program.
While this array is initially empty, it is filled with all the values read by the program from locations that have not been initialized before.
To keep track of the initialization, the \emph{Initialized} array records every byte that has been accessed, either read or written, at least once.
Writes modify the object array, which is the memory that is accessed, as well as the initialized array.
Reads take their value from the object array, except if the initialized array indicates the values have not been accessed before. 
In that case, the a value is generated and stored in the object array as well as the input shadow.
In any case, the initialized array will record that the memory region was accessed.

The example in \cref{fig:object_handling} illustrates the effects on the internal runtime state as memory is accessed.
First, the \texttt{WRITE 1} is performed which impacts the initialized array and the object array content.
A read of three prior untouched locations follows.
The runtime generates new values, here 2, 3, and 4, which are stored in the object array, but not the input shadow as the locations were not initialized before. Finally, the written locations are marked as initialized.
The subsequent write to a fresh, initialized location modifies the object array and the initialized state.
At this point the object array and the input shadow diverge.
Final reads from initialized locations are simply read from the current object array state.
As the example illustrates, the initialized, hence used, part of the object array can be larger than the used part of the input shadow.
Since replay only requires the initial values in the shadow, our stored input file will not include the user initialized parts of the objects.
However, when replaying the input, we still need to allocate memory for parts of the object that get accessed during the execution.
Consequently, we store metadata about how big the allocation needs to be and at what offset. Both can be inferred from the \emph{Initialized} array.

\begin{figure}[t]
\includegraphics[width=0.9\linewidth]{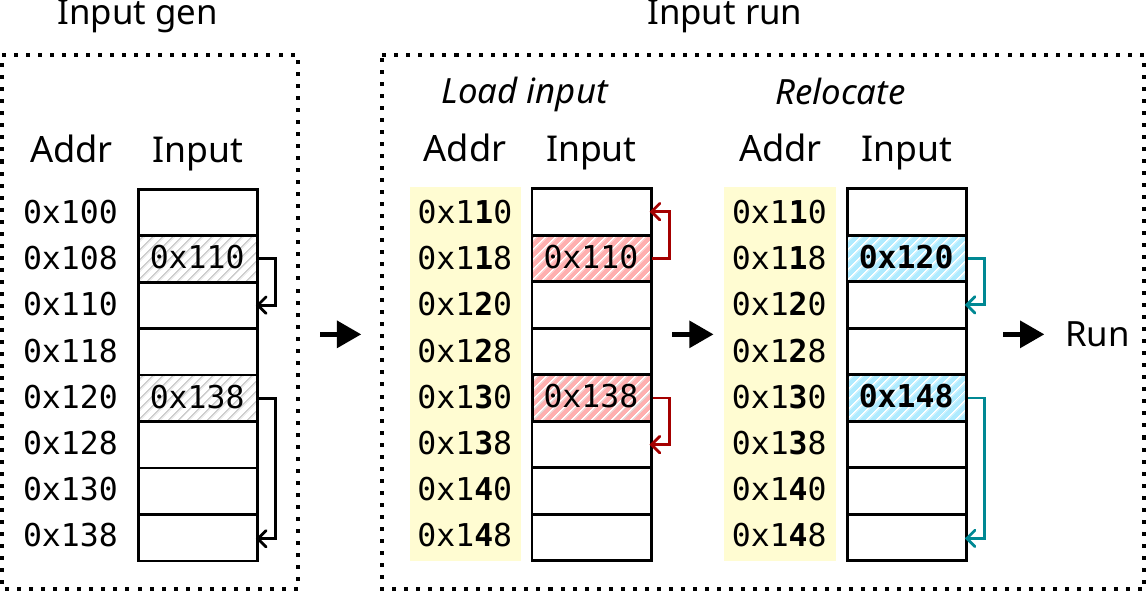}
\vspace{-2mm}
\caption{Pointers in inputs loaded from a file must be fixed to point to the correct location in the loaded input.
This is achieved by storing metadata about which locations in the input contains pointers (patterned background) and what object id and potential callee is at which location.}
\label{fig:pointer_relocation}
\end{figure}

\begin{figure*}[t]
    \centering
    \small
    \begin{tabular}{lrrrrrr}
    Language & Functions & Compiled & Inputs (normal exit) & Inputs (all) & Ran (normal exit) & Ran (all) \\
\hline
C &  667,519 &  658,884 (99\%) &  591,229 (89\%) &  607,748 (91\%) &  582,002 (87\%) &  598,600 (90\%) \\
C++ &  7,038,354 &  7,032,629 (100\%) &  6,333,772 (90\%) &  6,465,452 (92\%) &  6,187,666 (88\%) &  6,302,387 (90\%) \\
Julia &  3,731,668 &  3,731,668 (100\%) &  2,317,628 (62\%) &  3,500,173 (94\%) &  2,279,240 (61\%) &  3,423,596 (92\%) \\
Rust &  6,380,366 &  6,276,211 (98\%) &  5,236,692 (82\%) &  5,568,630 (87\%) &  5,123,473 (80\%) &  5,427,870 (85\%) \\
Swift &  6,058,725 &  6,050,627 (100\%) &  5,660,798 (93\%) &  5,738,468 (95\%) &  5,641,015 (93\%) &  5,720,465 (94\%) \\
\hline
Total &  23,876,632 &  23,750,019 (99\%) &  20,140,119 (84\%) &  21,880,471 (92\%) &  19,813,396 (83\%) &  21,472,918 (90\%)
    \end{tabular}
\vspace{-2mm}
\caption{The number of functions we were able to instrument, generate inputs for, and for which the generated inputs ran successfully. Generation was attempted five times with different seeds. We also keep track of whether the exit was abnormal or normal (see \cref{sec:program-exits}).}
\label{tab:all_language_results}
\vspace{-2mm}
\end{figure*}

\paragraph{Loading Input Files}
To replay recorded or generated inputs that feature pointers, one has to either place the memory at the same (virtual) address, or update the pointers to match the addresses in the run environment.
Related work~\cite{parasyris2023kernelrecordreplay} has described various problems with the former approach which is why we opted for the latter.
In general, it has less requirements on the operating system and readily allows the executable to be ported to different environments.
However, it forces the runtime to update all pointers, including function pointers, when loading a generated input since the allocations in the run environment do not match the addresses in the generation environment.
The problem is illustrated in \cref{fig:pointer_relocation}.
In the leftmost part, the memory state during recording, which is the state that is stored, is shown.
The generation runtime is aware of pointers into runtime-managed memory due to the type that was used to initialized the location, or generate the value (ref.~\cref{tbl:runtime_calls}).
In \cref{fig:pointer_relocation}, pointers are highlighted with a patterned background.
The middle parts shows the state once the replay runtime restores the input file at an address with 16 bytes offset.
Before replay execution of the function can begin, all pointers need to be reattached to their original target, as indicated in the right part.
This happens for both data pointers and function pointers stored in memory and present in the stream of generated values.
To allow such a rewrite, the runtime maps objects and potential callees to a numeric value during input generation.
The mapping, stored in the input file, is used at replay time together with a mapping of these indices to the new locations of the corresponding objects.
A pointer that points into object $K$ at offset $N$ in the input file is updated to point into the new location of object $K$, plus the offset of $N$, at replay time.

\paragraph{Aligning Objects}
As memory instructions can be sensitive to alignment, we ensure the replay state aligns the objects at least as much as the recording run did.
An example for this is shown in the lower part of \cref{fig:object_handling}.
The stored inputs, thus the values 2, 3, and 4, are placed in a larger object allocation that is big enough to hold all values stored in the object later.
This object allocation is then embedded in a padded allocation to ensure the alignment matches the generation run.

\begin{figure}[b]
\footnotesize
\begin{tabular}{l|rrr}
\centering
Lang. & \#\,Modules & Avg \#\,Funcs/Module & Avg \#\,BBs/Func \\
\midrule
C & 31610 & 20.84 & 16.40 \\
C++ & 10545 & 666.92 & 3.02 \\
Julia & 4918 & 758.78 & N/A \\
Rust & 11735 & 534.83 & 6.20 \\
Swift & 48972 & 123.55 & 4.08 \\
\end{tabular}
\vspace{-2mm}
\caption{Per language statistics of the modules from the ComPile dataset that were used for evaluation.
The basic block count for Julia is currently not reliable and omitted.}
\label{tab:dataset-statistics}
\vspace{-1mm}
\end{figure}

\section{Evaluation}
\label{sec:evaluation}

To evaluate the efficacy of our technique, we performed several large scale studies on approximately one quarter of the
LLVM-IR dataset ComPile \cite{grossman2023compile}.
We analyzed how many functions inputs were successfully generated, the percentage of basic blocks they execute, and the impact of multiple input generation passes combined with coverage-guided input generation.
\Cref{tab:dataset-statistics} shows that we are testing on approximately 100 thousand modules (translation units) that contain, on average, hundreds of functions each.
The evaluation was parallelized across modules using Jug \cite{coelho2017jug}.
We generated $105$ million inputs for $21.5$ million functions using $245$ node hours on 48-core Zen2 nodes,
which results in an approximate throughput of 2200 functions per core hour.
We limited the total parallelism per node to only utilize 40 cores due to memory constraints.
All experiments limited the time for input generation and replay to five seconds per input.

For each module, we attempt to generate inputs for every function definition.
The results in \cref{tab:all_language_results} shows the success rates at each stage of generation and replay.
We list inputs and runs that return to the entry point as \emph{normal exit} while \emph{all} include executions that end with an \texttt{ig\_exit} (ref.~\cref{sec:program-exits}).
However, these values are difficult
to interpret as many functions, e.g., exception handlers, do not have ``normal'' exit paths. 
We attempted generation of five inputs with different random seeds and a five second timeout each.
If any seed succeeded the function was counted as a success.
We achieve a nearly perfect ($99\%$) instrumentation and compilation rate due to module preparation (ref.~\cref{sec:preparation}).
Input generation terminates successfully for approximately $92\%$ of all functions.
Most of these, a total of $90\%$, can also be replayed successfully.


While function execution at scale is informative, it does not give any indication of function coverage.
Inputs that only exercise early exits are less interesting than inputs that cover many blocks in a function.
The compiler pass will introduce profile instrumentation, using LLVM's PGO infrastructure,
into the replay code which enables determination of the execution count of basic blocks at runtime.
This profile information is utilized by subsequent instrumentation runs, as described in \cref{sec:value-gen-hints}.
The profile information is further used to generate the basic block and coverage data presented here.

\begin{figure*}
\includegraphics[width=0.85\linewidth]{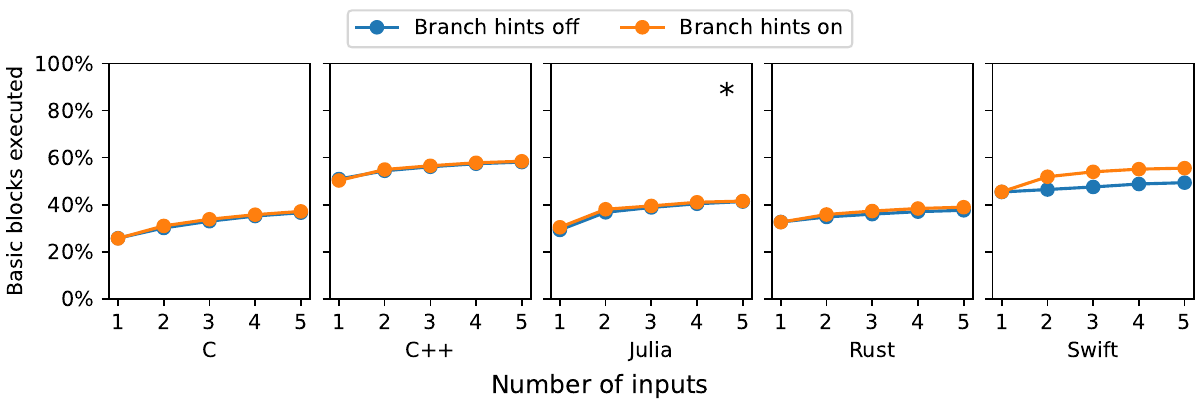}
\vspace{-2mm}
\caption{The average percentage of basic blocks executed in a module grows with the number of generated inputs for each function (* Our coverage data for Julia is not as reliable as we only have coverage statistics for 23\% of modules due to a issue with LLVM's PGO framework.)}
\label{fig:coverage}
\end{figure*}

Coverage results in \cref{fig:coverage} show that, on average, $37\%$ of all blocks are executed with a single generated input and coverage increases to $45\%$ when five generated inputs are used, indicating interesting inputs are generated that explore
different parts of the program.

\begin{figure*}
\includegraphics[width=1.0\linewidth]{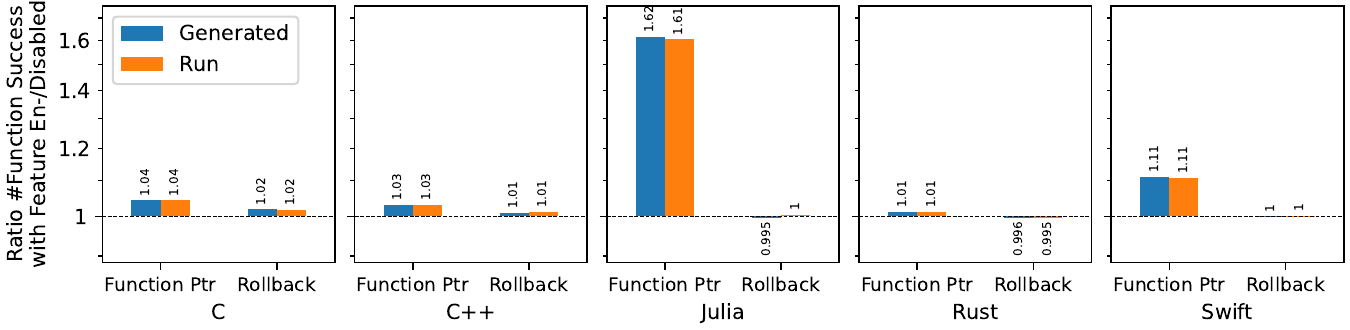}
\vspace{-7mm}
\caption{Ablation study. Each bar shows the relative change in how many functions we could generate inputs for or run with generated inputs upon enabling the feature. The two features we investigate are function pointer handling (\cref{sec:indirect_calls}) and rollback support for pointer generation (\cref{sec:rollback_offset}, \cref{sec:rollback_null}). Note that even 1\% improvement amounts to more than 200,000 functions.}
\label{fig:ablation}
\end{figure*}



\subsection{Ablation Study}
\label{sec:ablation-study}
We evaluated the impact of specific features on the success rate for generating and running inputs.
We have chosen to determine the effect of branch hints (ref.~\cref{sec:value-gen-hints}), function pointer handling (ref.~\cref{sec:indirect_calls}), and rollback support object comparisons (ref.~\cref{sec:rollback_offset}).

\paragraph{Branch Hints}
\cref{fig:coverage} shows the coverage information for both having branch hints turned on and off over five input generation runs.
The coverage generally increases with the number of runs and the increase is slightly higher when branch hints are enabled.
For Swift this feature is especially effective and the accumulated difference led to a 12\% coverage increase with five inputs.
For other languages the improvement is roughly 1\%.

\paragraph{Indirect Calls}
\cref{fig:ablation} shows the effect of disabling function pointer support and rollback across languages.
The former is especially important for Julia and Swift due to the amount of indirect calls.
With Julia, the number of functions for which valid inputs were generated dropped to 57\% without support for function pointers.
For Swift, there is still a notable improvement of about 11\% from adding explicit function pointer support.
This shows that the implementation and idioms of these two languages heavily depend on abstraction through indirect calls.
In the other three languages, we still see improvements.
However, Julia has one order of magnitude more indirect calls per function, namely 1.3, than any other language.
Swift features 0.3 indirect calls per function, followed by C with 0.2.
C++ and Rust are trailing with 0.04 and 0.02 respectively.

\paragraph{Rollback}
Rollback to support object comparisons is mostly impacting C and C++.
This might be explained by the use of pointer based iterators and null-pointer comparisons in these languages.
Since this feature introduces another source of chance and increased the code generation time, some runs might only succeed if it is disabled.
This explains the minimal regression for Rust.

\begin{figure}[ht]
\centering
\begin{minipage}{0.9\linewidth}
\vspace{0.2cm}
\begin{minted}[fontsize=\footnotesize, escapeinside=||]{cpp}
void gemm(int ni, int nj, int nk, double a,
          double *C, double *A, double *B) {
  int i, j, k;
  for (i = 0; i < ni; i++) {
    for (j = 0; j < nj; j++) {
      C[i * nj + j] = 0;
      for (k = 0; k < nk; k++) {
        C[i * nj + j] += a * A[i * nk + k]
                           * B[k * nj + j];
      }
    }
  }
}
\end{minted}
\centering
\textbf{(a) Code}
\vspace{0.2cm}
\end{minipage}\\
\centering
\begin{minipage}{0.9\linewidth}
\footnotesize
\begin{tabular}{lr|r|r}
 & Initialization & Run & Input Dump \\
\hline
Input Generation & 559 $\mu$s & 166,309 $\mu$s & 5,510 $\mu$s  \\
Input Run & 634 $\mu$s & 3,421 $\mu$s & \\
Original & 286 $\mu$s & 3,334 $\mu$s &  \\[0.2em]
\end{tabular}
\centering

\small
\textbf{(b) Runtimes}
\end{minipage}
\begin{minipage}[b][22mm][b]{0.35\linewidth}
\centering
\footnotesize
\begin{tabular}{lr}
 & Input \\
 \hline
 \texttt{ni} & 141 \\
 \texttt{nj} & 152 \\
 \texttt{nk} & 183 \\[0.2em]
\end{tabular}\\
\centering

\small
\textbf{(c) Generated Input}\\
\end{minipage}
\begin{minipage}[b][22mm][b]{0.6\linewidth}
\centering
\footnotesize
\vspace{0.2cm}
\begin{tabular}{lr}
 Required storage for \\ \texttt{A}, \texttt{B}, \texttt{ni}, \texttt{nj}, \texttt{nk}, and \texttt{a} & 428,972 bytes\\
 Input File Size & 429,380 bytes \\[0.2em]
\end{tabular} \\
\small
\textbf{(d) Generated Input Storage Size} \\
\end{minipage}
\caption{Generic matrix multiply \textbf{(a)} for which we generate inputs \textbf{(c)}. We time the various stages in our runtimes \textbf{(b)} to investigate what the overhead of our input generation bookkeeping is. We also evaluate how close to optimal our input storage is \textbf{(d)}.}
\label{fig:gemm}
\end{figure}

\subsection{Input Storage Footprint}
\label{sec:eval_storage_footprint}
To evaluate the overhead of our input files, compared to native inputs, we adapted a general matrix multiple code from Polybench~\cite{pouchet2012polybench}, shown in \cref{fig:gemm}.
Input generation resulted in matrices with the dimensions $141\times152\times183$ (ref.~\cref{fig:gemm} (c)).
Since the initial state of matrix \texttt{C} is not read, only matrix \texttt{A} and \texttt{B} are inputs.
To save the initial state of matrix \texttt{A} and \texttt{B}, as well as the values of the parameters \texttt{ni}, \texttt{nj}, \texttt{nk}, and \texttt{alpha}, one needs 428,972 bytes worth of storage.
Our runtime implicitly infers that the memory occupied by \texttt{C} is not an input that needs to be store in the input file.
However, the effective allocation size of \texttt{C}, all the aforementioned inputs, and metadata is stored in the input file.
Compared to the minimal required storage for these dimensions, our runtime adds an additional 408 bytes, an overhead of roughly $0.01\%$.
However, the overhead heavily depends on characteristics such as the number of objects, object sizes, number of pointers.

\subsection{Generation Bookkeeping Overhead}
\label{sec:eval_generation_overhead}
Using the same example, \cref{fig:gemm}, we evaluate what the runtime cost of our bookkeeping is.
On this specific benchmark, we find that the generated overhead is $50\times$ the native execution speed.
However, the time required to execute the kernel in our replay step is indistinguishably close to the original benchmark time.
That said, it takes $2.2\times$ more time to initialize the input arrays in the replay run compared to the original, native initialization.
As before, these numbers are heavily dependent on the code characteristics and the inputs.
Dense compute heavy codes, like this one, are especially hard for our scheme as the instrumentation prevents potent optimizations, such as vectorization.

\section{Future Directions}
\label{sec:future_work}



There are several directions that can help in achieving a higher coverage of both the number of generated inputs as well as coverage of executed code paths.
One example is stubbing system calls, which applications will sometimes use directly.
We currently do not handle these cases in any special way, and thus might be missing out on coverage.
In addition to stubbing additional elements to allow us to increase
coverage, taking advantage of existing research in feedback directed test generation
\cite{garg2013feedbackconcolic,pacheco2007feedbackrandomtest} and other techniques like static analysis for covering specific blocks 
\cite{garg2013feedbackconcolic,visser2004pathfinder,yoshida2017klover,ringer2017solverlanguage} might allow us to significantly increase
coverage.

Given that our primary focus of our tooling is large-scale performance analysis, it is prudent that the inputs that we
are generating exercise the code in the same way as it is in production systems. Given that our inputs are randomly
generated, this is unlikely to be the case. We leave it to future work to explore how well our inputs match the profiles
of the same code snippets used in production systems. We are also interested in exploring how we can guide
input generation using profiles collected from production to help match the code execution paths. Finally, we believe
it would be crucial to explore how important this aspect is for other work like ML-based performance prediction.

In addition to the listed additions to the tooling, we would also like to invest in upstreaming the tooling to the LLVM project
so that it is more broadly available to those interested in using it, easier to contribute to for those wishing
to make improvements, and better tested across multiple architectures.

On top of further improvements to the tooling, we believe that this tooling directly enables several exciting
research directions. Large datasets of executable code are currently quite sparse and existing approaches \cite{armengol2022exebench}
have so far resulted in relatively small datasets due to the use of small initial amounts of code in addition to low
generation and/or replay rates. Our relatively high replay rates, along with recent dataset advances allows for the creation
of datasets with significantly more executable functions. With the recent explosion
of ML focused on code, having more inputs to train ML to reason about code, including execution about code is becoming
increasingly important \cite{shi2020neuralcodefusion}, and large datasets are essential for achieving peak performance
\cite{hoffman2022chinchilla}. 


With a large dataset of runnable function-level snippets created through this tooling, large-scale performance introspection
also becomes possible. This might enable large-scale analyses of compiler runtime performance improvements, regression tracking,
and evaluations of various optimizations and hyperparameters. In addition, a variety of runnable snippets along with benchmarking
info would allow for the tuning and creation of better performance models, both classical and ML-based. While there are a variety
of existing performance models using a variety of techniques such as simulation \cite{binkert2011gem5}, analytical models
\cite{abel2022,abel2023facile,laukemann2018osaca}, and ML models 
\cite{li2022simnet,sykora2022granite,mendis2019ithemal,amalou2024transformerprediction},
these have often not been evaluated on large datasets
or have suffered from a lack of training data in the ML case. In addition to the lack of data in the ML case,
previous datasets such as BHive \cite{chen2019bhive} were often only at specific granularities,
like that of the basic block, that prevented modeling of performance at a broader level.

\section{Conclusions}
\label{sec:conclusion}

We present input-gen -- a tool that uses sanitizer-inspired compile-time instrumentation to generate stateful inputs with a high success rate.
We evaluate this tooling extensively across a large dataset to demonstrate efficacy.
We generate working inputs for $90\%$ of all functions, or $21.5$ million function in total.
After profile guided generation of five inputs per function, block coverage reached an average of 45\%.
While the current input generation execution time overhead is substantial, roughly $50x$, subsequent replaying of the inputs comes at close to no extra cost.

We improve significantly upon previous work in multiple dimensions.
Our approach is applicable across languages, allows direct and indirect function calls, is guided to correct past decisions in future runs, and produces stateful inputs for substantially more functions than any other work we could locate. 

We believe that the availability of an input generation tool that scales to massive datasets, together with the consequential datasets of inputs, will open up novel possibilities for ML-based techniques in compilers and find uses in testing, profiling, and benchmarking.

\section*{Acknowledgements}

The views and opinions of the authors do not necessarily reflect those of the U.S. government or Lawrence Livermore National Security, LLC neither of whom nor any of their employees make any endorsements, express or implied warranties or representations or assume any legal liability or responsibility for the accuracy, completeness, or usefulness of the information contained herein.
This work was in parts prepared by Lawrence Livermore National Laboratory under Contract DE-AC52-07NA27344 (LLNL-CONF-865462).

This work was supported by JST SPRING, Grant Number JPMJSP2106 and the RIKEN Junior Research Associate Program.


\bibliographystyle{ACM-Reference-Format}
\bibliography{references}

\end{document}